# *What is Osmosis? Explanation and Understanding of a Physical Phenomenon*

***Abstract***. *Although osmosis is a familiar phenomenon, and of pivotal importance in natural systems, it is seldom explained how it might work on the molecular level (if treated at all in physics text books). The standard treatment of osmosis in thermodynamics employs the concept of the chemical potential and does not give any clues how the process "really" works. On the other hand one may also encounter conflicting qualitative and molecular explanations of osmosis. We use the case of osmosis in the present paper to elucidate different ways to "explain" and "understand" physical phenomena. The role of qualitative understanding of physical phenomena is emphasized. However, as the case of osmosis demonstrates, there may be a big gap between the abstract macroscopic theory and molecular conceptions of the mechanism.*

Frank Borg ( borgbros@netti.fi ). Jyväskylä University, Chydenius Institute, PB 567, FIN-67101 Karleby.

# Table of Contents

## 1. Introduction

Of course, we all know what osmosis is. It is about membranes and concentration differences that makes water (considered in most cases) flow against the solute concentration gradient, a phenomenon which is typically explained in terms of the "osmotic pressure" given by the classical van't Hoff formula (1885),

(1)     $\Pi = kT\, c$

This relates "osmotic pressure" $\Pi$ for a solute to its concentration $c$, the number of solute particles per unit volume, when $c$ is "small" ($k$ is the Boltzmann constant and $T$ the absolute temperature). (That $c$ is "small"; i.e., we have a dilute solution, means that it is much smaller than the concentration of the solvent.) The formlula (1) predicts a 25 atm pressure for a solute concentration of 1 mole per liter (= 1M) at 20° C.

The traditional derivation of (1) is based on the use of the chemical potential. Now I believe that only a few exceptional students, when exposed to the traditional treatment, may get a grasp of what is "really" taking place in an osmotic process on the molecular level. A phrase like "the decrease of the chemical potential lowers the pressure" does perhaps not help the student to intuit a physical explanation of the process. Probably there will be left lingering a notion of some mystic "osmotic force" that pushes the molecules around. Thermodynamics provides very effective theoretical methods, but sometimes drawing results from manipulating the thermodynamic potentials may seem

Frank Borg ( borgbros@netti.fi ). Jyväskylä University, Chydenius Institute, PB 567, FIN-67101 Karleby.

more like magic (to me anyway) than giving a physical explanation for the phenomenon considered[1]. Basically, thermodynamics predicts the osmotic effect but does not "explain" it. Furthermore, classical thermodynamics deals with equilibrium states, not with the transition from one state to another.

The continental philosophical tradition has distinguished two basic modes of comprehending things: *understanding* (Verstehen) versus *explanation* (Erklären) (for an interesting account see von Wright 1971). Thus, "understanding" has been associated with humanistic inquires where one tries to understand motives and reasons, whereas explanations has been regarded as typical for the natural sciences looking only for cause and effect relations. Still, I think many physicists are also striving for an "understanding" of their subject matter; few are content with just formal deductions. In this case understanding is a question of relating the phenomenon to basic physical principles and models, or something that is already familiar from experience.

Two dimensions of a scientific field of inquiry can also associated with the *syntax part* and *semantic part* of a theory. The syntax part consists basically of the formal symbols, rules and "axioms", which as such is "empty" and gains meaning only through its semantic part which interprets the symbols and provides models that realizes the abstract formalism. Now, the "explanation" of osmosis given strictly in terms of the chemical potentials can be considered to be largely syntactical (consisting mainly of formal manipulations) and may as such provide little cues to the meaning of it. Neither does the thermodynamical derivation explain the osmotic effect in the sense of giving a causal chain showing how the effect is obtained.

Osmosis will be used here to illustrate the case where we have explanations and understandings of a phenomenon on different levels and of different kinds. We may have an explanation on the microscopic or macroscopic level; we might try to explain what happens in terms of molecular interactions, or only by giving an abstract mathematical derivation in terms of "thermodynamic potentials" that do not necessarily convey any idea of how the actual process takes place. *Of course, it is the very abstract and general nature of thermodynamics that makes its methods so powerful, but in the end we have also to try to understand the "mechanism" on the molecular level.*[2] Such considerations may suggest new effects not apparent given the general thermodynamic treatment. This consideration is also of pedagogical interest because relying only on abstract mathematical demonstrations enhances the danger that students memorizes formulas without gaining proper insight into the physics involved. Not least the growing interest in nanotechnology and soft matter physics will require some molecular considerations of the physical phenomena in an early stage of the teaching. Finally, when explaining physical issues to the public (popularizing science) we also need illuminating "pictures" of the physical mechanism. One could try to make it a rule to give a good verbal or "pictorial" physical explanation of a phenomenon before presenting the mathematical

---

1  For instance del Castillo (1997) acknowledges that despite the phenomenon of osmosis has been known for over 160 years its physical meaning remains obscure and that there is still no accepted theory of osmotic flow: "Por otra parte, a pesar de que el fenómeno de la ósmosis tiene 160 años de haber sido descubierto, su significado físico aún permanece oculto .... A partir de muy pocas observaciones se ha querido descubrir ese significado oculto, involucrando un enorme esfuerzo de comprensión intelectual, pero la idea verdadera no ha cobrado forma racional y actualmente no hay una teoría que explique y describa el flujo osmótico; aunque por el momento existe una teoría predictiva que ha hecho posible su aplicación práctica."

2  Barrow (1966 p. 643) states, after having presented the standard thermodynamical account for osmosis: "The expressions for the osmotic pressure derived in this section will apply as long as a membrane is available that will pass solvent and will not pass solute. The process by which it accomplishes this is immaterial". Indeed, as long as only the equilibrium states are of concern such an abstract description of the membrane may suffice, but as soon as we start to e.g. query about timescales of the process (does it take years to reach equilibrium?) the details of the mechanism become important.

Frank Borg ( borgbros@netti.fi ). Jyväskylä University, Chydenius Institute, PB 567, FIN-67101 Karleby.

formulation. Having an intuitive understanding of the mechanism will certainly also make the mathematical presentation easier to grasp. Not an uncommon situation is that we work out detailed and complicated solutions to problems but if asked we would not be able guess in which direction the process would proceed (the problem of getting the sign right - familiar?).

However, when we move to advanced theories like quantum field theory this approach may no longer be feasible (we may though presume that physicists working in the field employ various "pictures" of the processes in order to facilitate their understanding), but at this stage the student is much on his/er own already anyway. On the other hand it may be thought that such abstract inquires could be badly in need of good "pictures" to help intuit the content and give guidelines for further progress. I guess that one reason that some researchers are more successful than others in their field of study may be that they are guided by simple but effective "vernacular" models[3] and intuitions how things "really" work; we could perhaps speak of a certain sort of an intimate relationship where the researcher learns to interpret the cues given by Nature. The informal and personal nature of these "imageries" may make them difficult to pass over to a new generation of students, but at least there could more emphasis on developing an "intuitive" grasp and qualitative understanding of phenomena and their underlying theory. One way is it to train the ability to make rough back-on-the-envelope estimates of various quantities before attempting to give "complete" solutions to problems.

The discussion of osmosis shows that sometimes the verbal explanations may become somewhat muddled, e.g. due to unclear terminology, whence it may be impossible to assess how well the "explanation" accounts for the facts. The osmosis case also illustrates the phenomenon of a popular explanation - the "diffusion theory" - which is promulgated despite the lack of theoretical support.

As should be clear, what is said here is by no means a plea against abstract methods; on the contrary, I think that the abstract methods can be put to a more efficient use if we have a good qualitative understanding of the problem and its basic assumptions. As is also well known, "intuition" can sometimes lead astray; so, we need both to check each one another. Furthermore, it is not our suggestion that "understanding" necessarily requires a "mechanical model" for the phenomenon. E.g. contrived mechanical models of electromagnetic fields would hardly be of any use, pedagogical or otherwise[4]. Rather we would like to make a link to the honoured tradition of Gedanken-experiments where simplified models are devised such as to emphasize the essential elements behind a phenomenon.

---

3  One could also contemplate to model processes using macroscopic mechanical models. E.g. Janáček and Siegler (2000) present, with some reservations, a macroscopic toy modell for osmosis which they have tested in practice.

4  As is perhaps well known J Maxwell did indeed develop a "mechanical" model of the electromagnetic fields and waves as an aid in his comprehension of the subject. As he explained in a letter to Kelvin 1855 he tried to understand the subject "by the aid of any notions I could screw into my head". The modern *theoretical understanding* of electromagnetism is linked to the concept of the gauge field. Gauge theory provides a *structural framework* for understanding the mathematical form the theory takes. However, when a sicence is young we need sorts of Gedanken-experiments in order to probe the meaning of the constructs. Thus, much of the early discussion on the foundation of quantum mechanics was formulated using Gedanken-experiments (many of which have been realized in the laboratory since then). E.g. there was a debate on whether Heisenberg's use of the microscope Gedanken-experiment misrepresented the meaning of the momentum-position uncertainty relation.

Frank Borg ( borgbros@netti.fi ). Jyväskylä University, Chydenius Institute, PB 567, FIN-67101 Karleby.

## 2. Potential energy

Classical mechanics is largely based on using point-particle models and solving Newton's equation

(1)     $m_i \ddot{x}_i = F_i$

for each particle $i$ knowing the force $F_i$ acting on it. If we know the forces acting on the particles we can solve (in principle) the problem; this has become the "paradigm" of what it means to "explain" things in science. When we have a large number of particles we have to simplify the problem. If the particles form a "solid body" we can reduce the problem to one just involving the center of mass motion or rotations of the body. For deformable bodies and for fluids it is assumed that we can describe the motions using e.g. vector fields. We will discuss a simple problem which will later bring us to the osmosis problem. We have an U-shaped tube filled with water (see the figure). When the system is in equilibrium we assume that the water levels in both arms are on the same level $x_1$. Suppose then we have a situation where the water level on the right side has risen above the equilibrium level by $h$, and correspondingly been lowered by $h$ on the left side. Now we expect the water to flow so that the original equilibrium state will be restored. (We know there is friction so that the system cannot go on oscillating around the equilibrium for ever.) A traditional approach would be to calculate the potential energy $U$ of the system and determine its minimum which is expected to correspond to the equilibrium state of the system. Though this is quite a formal procedure it may provide a vivid picture of the system striving to reach the bottom of the potential well. Thus, in our case, if we assume that $h < x_1 - x_0$ we can disregard the potential energy of the water below $x_0$ because it does not change in the process. If the cross sectional area of the tube is taken to be $A$ we get for the potential energy of the water masses above $x_0$:

(2)     $U = \frac{1}{2} \rho g A (x_0 - h)^2 + \frac{1}{2} \rho g A (x_0 + h)^2 = \rho g A (x_0^2 + h^2)$

where $\rho$ is the density of water and $g$ is the acceleration of gravity. Expression (2) is obtained by summing the potential energy contributions $m g (x - x_0)$ of every "particle" above the reference level $x_0$. As we can see from (2) that $U$ indeed has a minimum at $h = 0$. If $M$ denotes the total mass of all the water we can also write the dynamic version (Newton's equation)

(3)     $M \ddot{h} = -\frac{\partial U}{\partial h} - \gamma \dot{h} = -2 \rho g A h - \gamma \dot{h}$

where we have introduced a frictional term $\gamma \dot{h}$ which makes the system come to rest asymptotically. Again, the stationary equilibrium $\dot{h} = 0$ is reached for $h = 0$.

Now, we can agree that the previous account gives a quite satisfactory solution to the problem as far as the assemblage of water molecules can be treated as constituting a macroscopic fluid with "frozen" degrees of freedom (here we only used one degree of freedom symbolized by the height $h$). Still we might be interested to know what happens on the molecular level. What makes the water molecules preferably flow in one direction instead of the other? Naturally, at this stage we do not ask for a detailed calculation of the forces and movements but for a qualitative understanding of the process at the molecular level not provided by the "potential formulation". Since water in this case can be regarded incompressible we know there must be strong repulsive forces between the water molecules when they come "too close". Thus to a certain approximation we may view the water molecules as impenetrable balls of a radius given by half of the typical minimum distance between two water molecules. With the water level risen at the right side we may expect on the average that a water "ball" on the right side experiences a higher "pressure" than a "ball" on the left side on the same level. The "balls" on the right side has to support a greater mass of water above them than the "balls" on the left side (on the same level) from which we conclude that the molecules will be pushed to the left and thus restore the equilibrium state. One could elaborate this story, but this is enough to make an interesting comparison with the osmotic case.

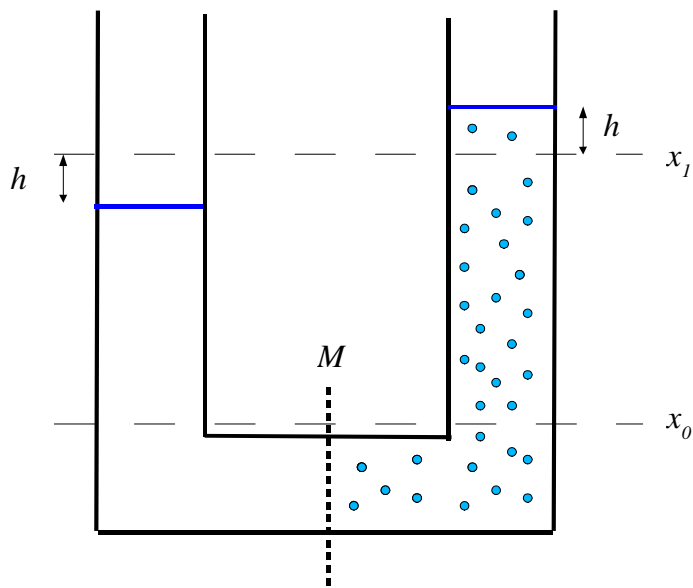

*Figure 1*

## 3. Osmotic pressure

Suppose we now insert a membrane *M* separating the U-tube in two sym-metrical halves as shown in the picture. Furthermore, we assume that water can freely pass through the membrane whereas molecules of the solute *S* cannot pass through it. We might e.g. imagine that the membrane contains holes small enough to let water molecules pass but too small for the solute molecules to pass (see the next figure)[5].

---

5  Biological membranes may have pores of diameter as small as 2 Å.

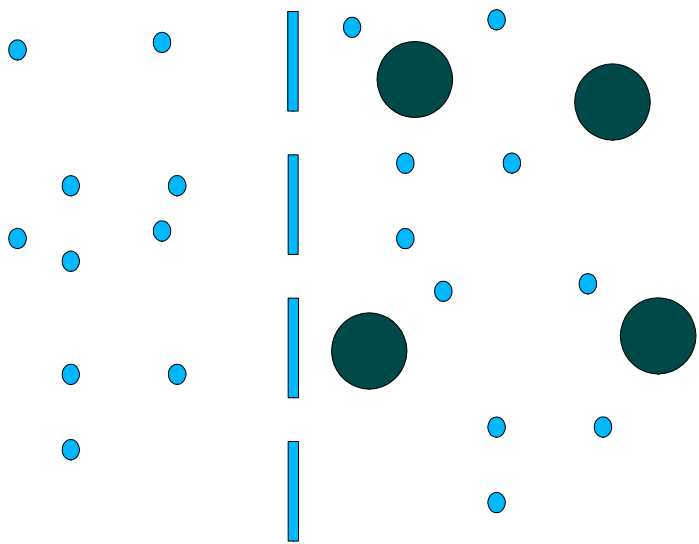

*Figure 2*

Thus, suppose that in the initial state the water levels are equal on both sides. Then we add a small amount of a solute to the right side and the water level is seen to start to rise - why? An important point is that osmosis is a *colligative property*: it does not depend on the specific nature of the solute (such as the exact size of the solute molecules), only on its concentration.

If you look into a traditional physical chemistry text you have to cover a quite a bit of ground to get the story. First you have to lay down the basics of thermodynamics, then you have to give an account of the Gibbs function and introduce the chemical potentials and finally apply these concepts to the problem at hand in order to derive (1), and in the end you will perhaps have no clue what really takes place. Also, the abstract theory may be consistent with a number of different, and even conflicting, molecular descriptions of the mechanism. The basic fact, however, is that the system is able to convert heat (the system is in thermal contact with the surrounding) into mechanical energy (water rising). The question is about the details of this perfomance.

### *4. Pressure and the virial theorem*

Before we proceed to discuss osmosis we have to discuss the concept of pressure. For gases the pressure at the boundary is easily understood as the result of molecules hitting the wall of the vessel. For an ideal gas the pressure is independent of the mass of the molecules and depends only on the average kinetic energy of the molecules which is given as *kT/2* per degree of freedom. For liquids we still have the same average kinetic energy per molecule, but the pressure (at the boundary) is much less than what one would obtain according to the ideal gas law. If the water exerted the same pressure as an ideal gas its pressure at room temperature and normal density would be around 1200 atmospheres. It is natural to assume that it is the attraction between the molecules which reduces the pressure. This is elegantly demonstrated mathematically by the *virial theorem,* which is a simple consequence of elementary mechanics[6] (derivation in the appendix),

(1) $$pV = \frac{2}{3}\langle E_{kin} \rangle + \frac{1}{6}\left\langle \sum_{i,j} F_{i,j} \cdot (r_i - r_j) \right\rangle$$

Here $\langle E_{kin} \rangle$ denotes the time average of kinetic energy, and $F_{ij}$ is the force by which the molecule at $r_j$ acts on the molecule at $r_i$. We obtain the ideal gas formula for noninteracting molecules (zero force) which have an average kinetic energy $3kT/2$. For attractive forces between the molecules the last term on the rhs becomes negative and the pressure is correspondingly reduced compared to the ideal gas case. Thus, the interaction term on the rhs may be viewed as a "negative pressure"[7] contribution canceling part of the kinetic pressure contribution due to thermal motion. If we increase the external pressure the last term on the rhs becomes less negative since the forces become more repulsive in order to hinder the molecules to collapse into each other.

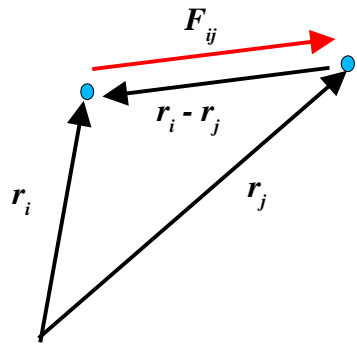

*Figure 3*

Consider, for simplicity, the zero gravity case where water is contained in a volume and forms a spherical mass in equilibrium with the vapour.

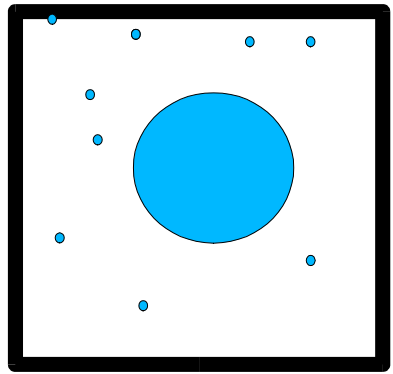

*Figure 4*

If we apply the virial theorem to the spherical mass of water the pressure on the lhs in (1) will refer to the vapour pressure. For room temperatures this is quite small; a larger part of the kinetic pressure is thus canceled by the *ww*-interaction term on the rhs. What happens if we add a solute to the water, say by a concentration corresponding to an osmotic pressure of 10 atmospheres? If we assume the solute molecules do not interact, then we may conclude, by applying the virial theorem to the solute molecules, that it is the water which "cancels" the extra 10 atmosphere pressure. Indeed, writing $p_{solute}$ for the solute pressure etc, and $s$ for the positions of the solute molecules, we get the for the solute virial equation,

$$(2) \quad p_{solute} V = \frac{2}{3} \langle E_{kin}^{solute} \rangle + \frac{1}{3} \left\langle \sum_n F_n^{(w)} \cdot s_n \right\rangle$$

---

6  See e.g. Goldstein (1980 §3-4). A somewhat more detailed account can be found in Landsberg (1990 §18.2).
7  This terminology makes sense since the force is attractive instead of repulsive.

($F^{(w)}$ stands for the total force of the water molecules acting on the solute molecule at $s_n$.) For a nonvolatile solute we may expect that the solute does not escape from the solution and the pressure on the rhs side may be put equal to zero. In the present case it means that the *ws*-interaction term on the rhs is equal to "-10 atm". Similarly we get for the water in the solution

$$(3) \quad p_{solvent} V = \frac{2}{3} \langle E_{kin}^{solvent} \rangle + \frac{1}{3} \left\langle \sum_i F_i^{(s)} \cdot r_i \right\rangle + \frac{1}{6} \left\langle \sum_{i,j} F_{i,j} \cdot (r_i - r_j) \right\rangle$$

($F^{(s)}_i$ is the force by which all the solute molecules act on the water molecule at $r_i$.) For homogenic distributions of the water molecules and the solute molecules in the volume we can argue that the *ws*-interaction term is of the same magnitude but of opposite sign compared to the value of the *ws*-term in the solute equation. Thus, in the present case we get in the solvent equation a "+10 atm" contribution in addition to the *ww*-interaction. Therefore, compared with the pure solvent case, the *ww*-term in the solution case has to cancel another "10 atm" which means that the water in the solution has negative pressure contribution "10 atm" below the value for a pure solvent. Pure water and the water in the solution are then essentially in different energetic states[8]. Connecting pure water and the solution using a semipermeable membrane creates an effective pressure difference between pure water and the water in the solution which amounts to the "osmotic pressure" given by $\pi_{osm} V = \frac{2}{3} \langle E_{kin}^{solute} \rangle$ . It could be put like this: *pressure exerted by the water molecules on one another in the solution is less (due to the solute intervention) than in pure water and this makes it possible for pure water to force itself into the solution till the ww-pressure difference is canceled*[9].

When later reading the paper by Lorentz (1892) I found his explanation to be much in the line with the presentation above though he uses very little mathematics and does not directly refer to the virial theorem[10]. I consider e.g. his equ (3) (Lorentz 1892) to be more or less a rendering of the virial theorem.

---

8  This provides a kinetic background for the thermodynamic analysis of osmosis in terms of the "chemical potential".
9  Patlak (www1) offers a nice Java-simulation where the water-solute inteaction is modelled such that 3 water molecules stick to every solute molecule. Of course this is a great oversimplification and also the captions may suggest that water concentration difference is the motive power of osmotic flow. In another version [Patlak www2] is more explicit about the role of the water-solute interaction as a reardation effect: "In other words, the chemical activity of water is inversely proportional to solute concentration! Understanding this 'retarding' effect of solute on solvent is the key to understanding osmosis."
10 Lorentz (1892) is indeed trying to make sense of osmosis directly in terms of molecular kinetics and without using involved mathematics. As Lorentz explains in the beginning of his paper: "Les lois relatives à la pression osmotique d'une solution étendue et aux phénomènes qui a'y rattachent sont si simples, qu'on est naturellement conduit à essayer de les déduire directement de la théorie cinétique, sans se servir de la thermodynamique". Sommerfeld (1952) after presenting the standard thermodynamic treatment of osmosis, and conceding to its "paradoxical" result, mentions indeed that H A Lorentz has provided a kinetic explanation. Prof. Claus Montonen (Helsinki univ.) pointed out the reference Lorentz (1892) to me, which is most likely the one that Sommerfeld was alluding to.

This broad explanation of osmosis using the virial theorem seems also closely related to the Hulett-Hammel-Scholander theory (HHS for short here[11]) among the various explanations offered in the litterature. HHS water tension can be interpreted as the *ww*-interaction term in the virial equation. The basic tenet of the HHS theory - that pure water under external pressure *p*, and solvent water under external pressure $p + \pi_{osm}$, are in equivalent thermodynamical states – makes sense considering the virial equation for the solvent water. Indeed, in the case discussed, if the external pressure for the solution is increased by "10 atm" then water has to respond such that the *ww*-term also increases by "10 atm". It thus reaches the same level as for the pure solvent whence the *ww*-pressure is the same for pure solvent as for the pressurized non pure solvent.

One obvious observation is that osmotic pressure can do work since it can e.g. raise water. An osmotic system can be likened with a heat engine which converts the heat of the environment (heat bath) into work. Indeed, the heat is transferred to the thermal motion of the solute molecules and the water molecules which are responsible for the pressure.

It is of interest in connection with the above method to discuss the case of a gas instead of a liquid solvent. As an example we may have $CO_2$ and a mixture of $H_2$ and $CO_2$, which are separated by a rubber membrane that is permeable only to $CO_2$. In this case, if we assume we can neglect the interactions between the gas molecules, we may expect that $CO_2$ will distribute itself so that at equilibrium there will be equal concentrations on both sides of the membrane. The mixture though will have a higher pressure because of the additional partial pressure by the hydrogen gas. In contrast to the liquid case, the pressure increase is not in this case due to the influx of the permeant through the membrane; the pressure would persist without the membrane. We also recognize that we might have cases that are somewhere between these extreme cases where "osmotic pressure" is due to both partial pressure and an influx.

## 5. Mechanistic theories

It seems that despite mechanical explanations for osmosis abound[12], the clarification that can be obtained applying the virial theorem has been generally overlooked. Still these various proposed mechanisms are of interest, since the virial theorem does not detail the exact mechanism, only what it will accomplish if satisfying some general conditions. Also, a detailed mechanical picture may suggest where to look for deviations from the ideal case which is the basis of the van't Hoff formula. We will therefore take a look at a selection of mechanisms for osmosis that have been proposed or that we invent here for illustrative purposes.

### 5.1. "Solute bombardment theory"

A "classical proof" of equ (1) employs a setup with a movable membrane. This time the tube is filled completely with water and sealed at both ends. When we add an impermeant solute to the right side of the system it is no longer in equilibrium since the solute molecules is supposed to "push" the membrane to the left. In order to maintain equilibrium an opposing force corresponding to the osmotic pressure has to be applied to the membrane. If we just add a solute to a bottle of water no pressure build up takes place, something this bombardment theory cannot explain. It is only when we try to separate the solute and the water using a semipermeable membrane that an effective "osmotic pressure" is felt[13]. The osmotic pressure of such a solution calculated using (1) (taking into account that NaCl dissociates into ions in water; i.e., the ion concentration is twice the given NaCl-concentration) at 27° C becomes around 7.4 atmospheres. The bombardment theory would make sense for gases if we may imagine that there is some sort of a lid put over the "solution". Thus, the theory may be applicable in a case where the solute interacts very weakly with the solvent. Clearly the similarity of the van't Hoff equation (1-1) has suggested to many versions of the bombardment theory. As Barrow (1966 p. 643) remarks:

> "This similarity of the expression to the ideal gas law led van't Hoff and others to some not very fruitful ideas that view the osmotic pressure as arising from a molecular bombardment process."

---

11 In a letter Hammel suggests the theory be stated as the "Hulett-Dixon-Herzfeld-Duclaux-Myselfs-Scholander-Hammel" theory in recognition of the main contributors; so, consider HHS a convenient abreviation.
12 Weiss (1996 p. 226) refers to a study by Guell (1991) which lists no less than 14 different proposed mechanisms for osmosis.

One point though is that ideas can be fruitful despite being wrong - e.g. the bombardment theory does suggest the correct formula (1-1) and the explanation also holds in the case of gasous mixtures.

## *5.2. Surface theory*

Since water is free too pass through the membrane we may expect the same concentration of water molecules per unit volume on each side, so that the hydrostatic pressure exerted by the water molecules on both sides will be the same at the beginning. However, on the right side we have the additional bounces from the solute molecules. These molecules will try to escape via the water surface. As they hit the water molecules in the surface layer they will get a kick upward and drag with them neighbouring water molecules. Therefore we might expect a net force on the water molecules that makes water flow to the right till the force is balanced by the weight of the risen water. Since the solute is of low concentration we may assume there is no interaction between them in the solution, and can thus be regarded as an ideal gas when it comes to calculate the pressure exerted by the solute molecules. Thus applying the "gas law" $pV = nkT$ ($n$ = number of solute particles) to this case we obtain the equation (1) for the "osmotic pressure". In fact, this "theory" has some similarities with one that has over the years been advanced by H T Hammel ("water tension theory", 1976[14]) in that it presupposes a cohesion, and there appears to be an obvious way to test it. Indeed, according to the proposed mechanism, if we inject the solute (e.g. dropping a crystal of the solute) near the surface we should see the osmotic effect commence earlier than when injecting it near the membrane. An experiment by Mauro (1979) has been cited as showing the opposite to be the fact; thus, it looks as if we should search for the "osmotic mechanism" at the membrane. Hammel's theory, though, makes sense if we link the concept of "negative pressure" with the water-solute interaction that cancels the kinetic pressure of the solute as described in the discussion of the virial theorem above.

---

13  It may be pointed out that Frazer and coworkers measured osmotic pressure up to 273 atmospheres in their experiments 1916-1921.
14  This theory goes back to Hulett (1903). Hammel is a strong supporter of the water tension theory as indicated by the following testimony: "I have had engraved on my gravestone the following declaration: 'A physiologist who measured xylem and phloem sap pressures in trees; who embraced Hulett's theory of osmosis and who recognized the diffusion of bicarbonate ions as the principal osmotic effect in Starling's hypothesis.' When Scholander and I finished our monograph on 'Osmosis and Tensile Solvent' in 1976, I said to him, 'Pete, you will never know the day when our ideas about osmosis become widely accepted.' Twenty-five years later, I repeat the same statement, now applied to myself. I feel obliged, therefore, to engrave our views in granite." ("News From Senior Physiologists", www.the-aps.org/publications/journals/tphys/2001html/October01/srphys.htm .) Indeed, though more than 100 years of research, the mechanism of the ascent of sap in plants cannot be said to be fully understood yet - see e.g. Tyree (1997).

### 5.3. Solute attraction theory

Suppose there is an attractive force between the solute molecules and water molecules. Thus, when the solute is near a pore it can "suck" in water from the pore into the solution. This suction explains the "negative pressure" felt by water and which forces it to flow to the solution side. A weaker interaction may be thought to be compensated by a less impeded flow of the solute molecules and thus the water molecules it drags with it. Also closely related to this theory is the "water thirst" idea by Gerald Pollack.

### 5.4. "Water thirst theory"

This idea is based on the hydration phenomenon. The solute molecule is thought to be surrounded by layers of water molecules that attract further water molecules. In Pollack's own words[15]: "If all hydrophilic or charged solutes attract comparably large water layers, effectively they have enormous 'thirst'. Now, put these solutes in a chamber surrounded by a semipermeable membrane, with water alone on the other side of the membrane. If there is not enough water in the solute-containing chamber to satisfy the molecular thirst, water will be drawn from the neighboring container. That's it. The force of osmosis is one of hydration, and the water needs to be supplied from anywhere it is available."

Related to the "solute attraction theory" and "thirst theory" is a modell in which the membrane is described as a potential $\phi$ acting only on the solute. Joos presents an argument in (Joos 1986) based on this idea whose basic elements also figure in (Villars & Benedek 1974) and (Guell 1991). Suppose the membrane is given as the plane $x = 0$ and that the solution is on the side $x > 0$. Joos argues that as the solute is repelled by the membrane "viscosity causes some of the solvent to be dragged along ... The liquid thus streaming away from the membrane is replaced by fresh solvent which passes through the membrane, for no barrier exists as far as the solvent is concerned. The solvent streams in until the pressure excess equals the osmotic pressure." One assumes thus that two balancing forces are acting on the solute: the potential $\phi$ and a pressure $p$ (this will be criticized below). Joos assumes further that at the equilibrium the solute density (particles per unit volume) follows the Boltzmann distribution

$$(1) \qquad c(x) = c_\infty e^{-\frac{\phi(x)}{kT}}$$

The potential is $+\infty$ at $x = 0$ and drops to 0 as $x$ increases. Equating the hydrostatic volume force and the membrane repulsion one obtains

$$(2) \qquad -\frac{\partial p}{\partial x} = -c\frac{\partial \phi}{\partial x}$$

Combining (2) with (1) it follows that

---

15  Personal communication (email 11. Dec 2002).

(3)  $\dfrac{\partial p}{\partial x} = -kT \dfrac{\partial c}{\partial x}$

Integrating (3) from $x = 0$ to $+\infty$ we obtain the van't Hoff formula in the dilute limit if $p(0) - p(+\infty)$ is interpreted as the osmotic pressure $\pi_{osm}$. Equ (1) can be justified by assuming a stationary state where the total flow of solute, which is a sum of one part due to the membrane repulsion and one part due to diffusion, is zero[16],

(4)  $0 = J_{solute} = -\dfrac{1}{\gamma} c \dfrac{\partial \phi}{\partial x} - \dfrac{kT}{\gamma} \dfrac{\partial c}{\partial x}$

If we reckon with the Joos' drag effect then the water flow into the solution side could be expected to be described by an equation of the form (the quantities are evaluated close to the membrane)

(5)  $J_{solvent} = \alpha J_{solute} + L(\pi_{osm} - \Delta p)$

The first term on the rhs describes the flow of water induced by the "solute drag", the second term (which can be argued for on the basis of the virial theorem - see above) describes the flow due to osmotic pressure and the hydrostatic pressure difference $\Delta p$ over the membrane ($L$ is a parameter which describes how well the membrane "conducts" water). From this it is apparent that the "drag term" is not necessary for explaining the osmotic flow. Indeed, we may assume that the solute flow drops to zero since no new solute enters the solution. The stationary state $J_{solvent} = 0$ of (5) is thus given by $\Delta p = \pi_{osm}$. Looking back we realize that Joos has given no arguments supporting the assumption behind equ (2). Clearly, there is no need to assume a "pressure force" balancing the flow of the solute since we may suppose that we have a redistribution of the solute till the state (1) is achieved. However, it is possible to formulate an approach based on flow equations of the form (5). This is exactly what is done in theories[17] which employ the theory of non-equilibrium thermodynamics[18] developed by Lars Onsager (1903-1976) leading to the so called *Kedem-Katchalsky equations*. Still, as a phenomenological theory the *Onsager-theory* does not reveal the mechanism of the process (see the Appendix).

### 5.5. *Water concentration-diffusion theory*

One of the most popular "folklore theories" of osmosis assumes that water flows to the solution side via "diffusion"[19] driven by the difference in the concentration of water molecules which is supposed to be lower in the solution than in the pure water. The following quotation from a chemistry text can be taken as representativ for this "theory":

---

16  The force of friction is asumed to be $-\gamma \dot{x}$ with the friction coefficient $\gamma$ related to the diffusion constant $D$ by $D = kT/\gamma$ ("Einstein-relation").
17  See e.g. Janáček and Siegler (2000).
18  For a classic textbook on the subject see Groor and Mazur (1984). Sommerfeld (1952) contains a clear and concise introduction to irreversible thermodyanmics in section 21. The theory has been adapted for biophysical applications by A Katchalsky and P F Curran in *Nonequilibrium Thermodynamics in Biophysics* (Harvard UP 1965). The Kedem-Katchalsky equations are also discussed by Weiss (1996 § 5.5).
19  "Selfdiffusion" is a name used for this.

> "Water molecules can pass through the membranes in either direction, and they do. But because the concentration of water molecules is greater in the pure water than in the solution, there is a net flow from the pure water into the solution."[20]

Another text presents the idea in a little more detail:

> "Water molecules are continually colliding with the membrane from both sides. However, the rate of collision is somewhat less on the solution side (right) than on the pure water side (left) because the concentration of water molecules in a solution is less than their concentration in pure water (...) Since there are more collisions from the left to than from the right, more water molecules pass through from left to right than from right to left."[21]

A popular text on physiology states that:

> "Because dissolved solute molecules occupy space that would otherwise be taken up by watermolecules, the higher the solute concentration, the lower the water concentration. As a result, water molecules tend to diffuse across a membrane toward the solution containing the higher solute concentration, because this movement is down the concentration gradient for water."[22]

Wolfe (2001) gives a somewhat more sophisticated argument, involving also the Boltzmann factor, but still the water concentration difference is seen as the primary mover:

> "Hydraulic equilibrium is achieved when the lower concentration of water in the high pressure side is balanced by the higher pressure and therefore higher energy $pV$ on that side".

The concentration difference may however be considered as a secondary factor for the osmotic process for liquids[23]. Apparently, the concentration-diffusion theories of osmosis stay around because they appeal to imaging the liquid as a gas, which of course cannot square with all the facts about liquids as was made evident in the section on the virial theorem. For instance, consider a closed bottle half filled with water at room temperature. If we follow the lead of the concentration-fiffusion theory we would expect water to flow into the empty (save vapour) half of the bottle since water concentration is much smaller there. Naturally, this does not happen because of the interaction between the water molecules. Another problem with the conentration-diffusion theory is that the osmotic flow is generally observed to be *convective* rather than diffusive. Indeed, diffusive flow is due to random motions and is quite different from a forced flow due to a pressure difference. A further argument against concentration-diffusion theory (i.e. water concentration gradient as a motive power) is that we have solute effects related to osmosis such as lowering of the freezing point, and the rise of the boiling point, which apparently have little to do with diffusion but are related to the solute-solvent interaction. The question of water transport through the membrane itself

---

20  Petrucci et al. (2002 p. 554).
21  Brescia et al. (1975 p. 186-7).
22  Martini (2001 p. 73).
23  There is of course a small variation in water concentration e.g. due compressibility; an increase of external pressure from 1 atm to 10 atm (20° C) increases water density by a factor of 1.00046. Also, as pointed out by Hammel (2002), for some solutions, like NaF dissolved in water, the water concentration (molecules per unit volume) may in fact be *higher* than for pure water.

Frank Borg ( borgbros@netti.fi ). Jyväskylä University, Chydenius Institute, PB 567, FIN-67101 Karleby.

is a different issue. E.g. for very small pores we can no longer describe the transport as a Poiseuille flow. Indeed, as small pores approaches a 1-dimensional system the traditional concept of diffusion looses its meaning.

## 5.6 Obstruction mechanism

Conisidering the sieve modell of the membrane one might hypothestize that osmotic effect arises from the mechanical obstruction to water flow from the solution due to the presence of the solute molecules. This mehcanical modell is e.g. found in Atkins and Beran (1992 p. 432):

> "The [osmotic] flow occurs because the solvent molecules can pass readily through the membrane from the pure solvent into the solution, but the presence of the solute molecules blocks the return of some solvent molecules from the solution into the pure solvent. As a result, the flow from pure solvent into the solution is faster than the return flow."

This explanation makes certainly sense, yet it cannot be the essence of osmosis. Consider the case[24] of a membrane made of a palladium foil which is permeable for $H_2$ but not for $N_2$ and other gases. It seems that the membrane functions as a catalyst; it does not let through $H_2$ directly but decomposes it into atoms that can pass through the membrane and runite on the other side. Moreover, the obstruction modell directly contradicts the case with gases since here we obtain equal concentrations of the permeant on both sides, the "osmotic pressure" resulting from the extra partial pressure of the non-permeant gas.

## 6. Some consequences

One very important application of the osmotic effect involves the living cells. Since the (animal) cell wall is not believed to be able to withstand[25] a significant pressure difference the cell must regulate the intracellular particle concentration so as to match the outside concentration, otherwise the cell either fatally bursts or dries out and shrinks. This explains why it is not advisable to drink e.g. distilled water ("hypotonic" fluid). An "isotonic" solution is obtained by adding about 0.15 mol/L NaCl to water. This corresponds to the intra- and extracellular concentrations for most vertebrates. In diabetics dehydration due to excessive glucose level in the blood stream (while the body tries to get rid of the excess glucose water follows with it) can cause a serious state called *hyperosmolar coma*. This high osmolarity of blood can damage the cells and may be letal if rehydration is not quickly administered[26]. Dehydration also occures when sweating. Sweat is a hypotonic solution so that in sweating the salt concentration of the extracellular fluid (ECF) *increases*. Thus the popular custom to take salt pills when exercising have no physiological basis and no apparent benefits have been demonstrated. "Body reserves of electrolytes are sufficient to tolerate extended periods of strenous activity, and problems with Na+ balance are extremely unlikely exept during ultramarathons or other activities that involve maximal exertion for more than 6 hours", as pointed out by Martini (2001 p. 994). From the evolutionary point of view it could be argued that individuals who loose too much Na+ during exertion would have been disadvantaged (salt was not readily available in prehistoric times) and the genetic pool has therefore drifted toward a more sustainable physiology in this respect. Martini also makes the remark that sport beverages may in fact cause cramps, diarrhea and other problems if they have a too high sugar concentration (over 10 g/dl).

---

24　Discussed e.g. by Barrow (1966 p. 643).
25　Plant cells however may have strong walls and osmosis plays here an important role for the *turgocity* (rigidity) of the plant.
26　In physiology one uses also the expression "oncotic pressure" meaning "colloid osmotic pressure of body fluids".

The picture of the "sieve"-membrane above immediately also suggests the *reverse osmotic* (RO) effect: If we e.g. use a piston on the right side to exert an extra pressure above the atmospheric + osmotic pressure water will flow to the left leaving the solutes behind on the right side; thus we can get cleaned water on the left. A practical application of this is the desalination and purification of water and to concentrate apple juice. The purification method using reverse osmosis is also being referred to as *hyperfiltration*. Hammel (2002) describes an experiment where one removes water from the pure water side in the osmometer such that the increased hydrostatic pressure difference overcomes the osmotic pressure and the water starts to flow from the solution; i.e. we get reverse osmosis. An important effect in this case is that the solute which flows with the water stops at the membrane and a solute concentration builds up there. The solute drag brakes the water flow with a force that may be many times greater than the flow resistance offered by the membrane (see further *Appendix B*). Practical hyperfiltration devices are constructed[27] such that the solution flows parallel to the membrane and carries away solute that otherwise would accumulate at the membrane.

Another consequence of adding the solute to water is that the boiling point increases. This can understood in terms of a reduction of the vapour pressure. If we consider (see e.g. Ma 1985 p. 326) the arrangement[28] of figure 3 where we have a closed O-tube, we realize that the water vapor pressure on the left side (pure water) must be greater than on the right side (solution) by an amount equivalent[29] to the hydrostatic pressure of a column of vapour of the height 2$h$; i.e.,

$$\Delta p_{vapour} = -p_{osmosis} \cdot \frac{\rho_{vapour}}{\rho_{water}}$$

If we assume that the vapour satisfies the ideal gas law we can write the above result as

$$\frac{\Delta p_{vapour}}{p_{vapour}} = -\frac{c_{solute}}{c_{water}}$$

Thus, adding a solute to water decreases the vapour pressure in proportion to the solute concentration - a result discovered empirically by Raoult in 1886 and which is often used the other way around in order to derived the van't Hoff formula (1-1).

---

27 For some practical and theoretical details about RO see e.g. Lachish (www2003) and EPA (1996).
28 Variation of a Gedankenexperiment used e.g. by Hulett.
29 This argument assumes that that the solute cannot leave the solution; i.e. it is nonvolatile.

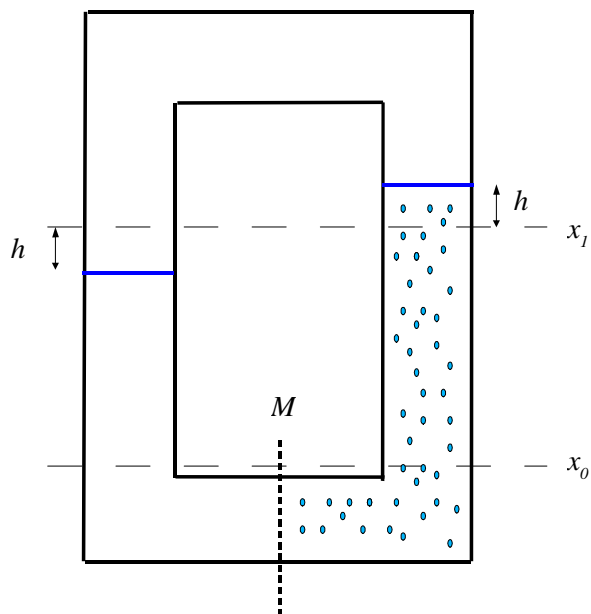

*Figure 5*

## 7. Historical note

A paper by Robert Boyle (1627-91) (Boyle 1682) could be one of the first studies of osmosis. Still, as the "father" of osmosis one may regard René Dutrochet (1776 - 1847) who studied osmosis (1828) using animal bladders (Dutrochet seems also to be the one who introduced the word *osmosis* in this context - greek for "push"). The botanists Wilhelm Pfeffer (1845 - 1920) and Hugo de Vries studied osmosis in plant cells. Pfeffer's classical setup (fig. 4) consisted of water (*W*), a glucose solution (*S*), and a semipermeable membrane (*M*) (the vessel containing the solute was made of nonglazed porcelain lined with ferro-cyan-copper $CuFe(CN)_6$ which made it permeable for water but not for glucose). van't Hoff (1852 - 1911) derived the formula (1) in 1885 using thermodynamic arguments[30].

---

30 Already back then osmosis became surrounded by disputes which at one time prompted van't Hoff to make the remark (1892): "Again we have the basically pointless question: What exerts osmotic pressure? Really, as already emphasized, I am concerned in the end only with its magnitude; since it has proved to be equal to the gas pressure one tends to think that it comes about by a similar mechanism as with gases. Let he, however, who is led down the false path by this rather quit worrying about the mechanism." (Quoted by Weiss (1996 p 185). A livley discussion of the topic has taken place in the pages of *American Journal of Physiology*, vol. 237, in 1979.

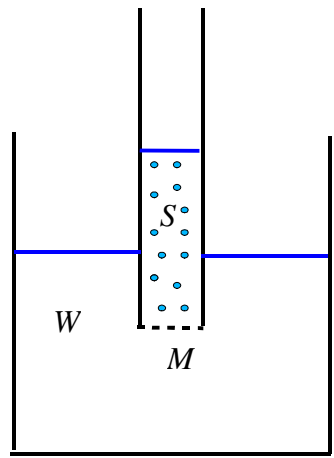

*Figure 6*

## 8. Thermodynamic explanation

We also wish to give an explanation of the osmotic pressure that uses the thermodynamic concepts but what we believe in a more transparent manner. First we might follow the lead by which we arrived at the potential $U$ in (2). It seems we only have to add the potential energy due to the solute. Apparently this would not predict the osmotic effect. As a comparison, minimizing the gravitational potential for a gas in a container we would predict that the molecules would lie dead flat on the bottom of the container. Well, this could be true in the exceptional case of a temperature of absolute zero. Normally temperature does not enter problems in mechanics but here is it apparent that it must play some role. The molecules do not grind to halt; in the collisions the gain and loss of energy averages to zero. We can no longer treat the solute or the gas as being fully characterized only by their gravitational potential energy. In thermodynamics this is expressed by the statement that we no longer have to minimize the potential energy $U$ but the combination

(1)     $F = U - T S$

usually referred to as "free energy". Here $U$ stands generally for "internal energy", $T$ for the absolute temperature and $S$ is entropy. How can it be justified to replace $U$ by $F$? We realize that for the solute at low concentration the pressure exerted by its molecules bouncing on and off greatly exceeds its "hydrostatic pressure" which is due to its weight alone. So one way out to proceed is to treat the solute in the "thermodynamic regime". A basic postulate of thermodynamics is that equilibrium is reached when the *total entropy* is a maximum. In special cases this can be translated into the requirement that the free energy $F$ be a minimum at the equilibrium. Therefore, free energy adds nothing new in content besides the entropy function, it is just a convenient way of determining the state of maximum total entropy under certain special conditions (constant volume processes). Indeed, let $S$ denote the entropy of a system, $S^e$ that of the surrounding, and $S^{tot}$ the total entropy. Suppose the system is in thermal contact with the surrounding. According to the 1st law of thermodynamics heat transferred from the surrounding to the system (with internal energy $U$, pressure $p$, volume $V$) is given by

(2)     $\Delta Q = \Delta U + p \Delta V$

Since the heat (2) is flowing from the surrounding (causing no irreversible changes in the environment) its change in entropy will be, according to the thermodynamic definition of entropy,

$$(3) \quad \Delta S^e = \frac{-\Delta Q}{T} = -\frac{\Delta U + p\Delta V}{T}$$

Now, according to the 2nd law of thermodynamics the total entropy cannot decrease,

$$(4) \quad \Delta S^{tot} = \Delta S + \Delta S^e \geq 0$$

Thus, for a system at constant *pressure* and temperature this is the same as

$$(5) \quad \begin{array}{l} \Delta G \leq 0 \text{ with} \\ G = U + pV - TS \text{ (Gibbs energy)} \end{array}$$

and for a system at constant *volume* and temperature we get

$$(6) \quad \begin{array}{l} \Delta F \leq 0 \text{ with} \\ F = U - TS \text{ (free energy)} \end{array}$$

How to apply these concepts to the present problem? If we expect the total volume to remain unchanged we may use the free energy $F$ in this case. The entropy of the water does not change when shifting its position in a constant gravitational field (entropy is only a function of the thermodynamic variables, such as pressure and the temperature). Here the free energy of water contributes thus only with the gravitational potential energy part $U$ to (1). Since we are considering the process to be isothermal, the internal energy of the solute may be taken as constant and does not affect the minimization problem. It remains to compute the entropy $S$ of the solute. Suppose the volume occupied by the solvent is $V$, and the solute can be treated as a gas in a volume $V$ in the sense that the solute molecules do not interact, then for its entropy we can use the expression for the ideal gas

$$(7) \quad S = S_0 + kn\ln(V)$$

where $S_0$ is a term not dependent on $V$ and $n$ is the number of solute molecules. Alternatively we could start with Boltzmann's definition of entropy,

$$(8) \quad S = k\ln(W)$$

where $W$ is the "thermodynamic probability" of the (macro)state; that is, the number of microstates consistent with the given macrostate. When the volume available for the $n$ particles of the solute increases from $V_0$ to $V$ it may be argued that the number of microstates (the temperature staying the same) for the given macrostate increases with a factor of $\left(\frac{V}{V_0}\right)^n$ from which we can again obtain the expression (7). Of course, if we use (8) we have to show how it relates to the thermodynamic definition of entropy used in (3). Next we can express the volume $V$ as

$$V(h) = V_0 + A h$$

and if we suppose $Ah/V_0 \ll 1$ and use $\ln(1+x) \approx x$ for small $x$, then we finally may set, disregarding unimportant constant terms, the free energy to be

$$(9) \quad F = \rho A g h^2 - T k A h \frac{n}{V_0}$$

We find that $h = 0$ is no longer a minimum for $F$. Instead it assumes minimum at a value for which

$$(10) \quad \Delta p = \rho 2 h g = k T \left(\frac{n}{V_0}\right)$$

The "pressure" $\Delta p$ corresponds to the hydrostatic pressure exerted by a mass of water of height $2h$. The new equilibrium state can now be interpreted as caused by the "osmotic pressure" exerted by the solute. It could be argued that this result is already contained in the "ideal gas" assumption (7) so that the next steps add nothing new to what we already knew form the ideal gas law. However, the method can be generalized so as to incorporate corrections to the ideal gas assumption. One of the simplest forms of corrections is taking into account the finite volume of the solute molecules. This will affect the entropy factor $S$ in (1). Indeed, calculating (7) we have to replace the factor $\left(\frac{V}{V_0}\right)^n$ in $W$ by

$$(11) \quad W = \prod_{i=0}^{i=n-1} \left(\frac{V - i v}{V_0 - i v}\right)$$

where $v$ is the volume excluded by one single molecule of the solute. Thus, the first solute molecule added to the solvent can occupy a space $V$, the 2$^{nd}$ one can occupy a space $V - v$ and so on. If the ratio $v/V$ is very small we get from (11) an approximate correction factor,

$$(12) \quad 1 - \frac{v}{V} \sum_{i=0}^{n-1} i + \frac{v}{V_0} \sum_{i=0}^{n-1} i \approx 1 - \frac{v}{2V} n^2 + \frac{v}{2V_0} n^2$$

which changes the entropy by an amount

$$(13) \quad k \ln\left(1 - \frac{v}{2V} n^2 + \frac{v}{2V_0} n^2\right) \approx -k \frac{v}{2V} n^2 + k \frac{v}{2V_0} n^2$$

Thus we get a second order correction to the "osmotic pressure" (10) given by

$$(14) \quad \Delta p = k T \left\{\frac{n}{V_0} + \frac{v}{2}\left(\frac{n}{V}\right)^2\right\}$$

In thermodynamic terms the "explanation" of the osmosis is that by adding a solute a new state of maximum total entropy is reached by water flowing into the solute side; the decrease in the total entropy due the heat flowing from the surrounding to the system (heat is "concentrated") is offset by an increase of configurational entropy of the solute. Using the thermodynamic free energy *F* this means that the osmotic equilibrium minimizes *F* for the water-membrane-solution-system.

## *8.1 Thermodynamic explanation - second version*

It might be of interest to look at a bit different version of the thermodynamic argument. Suppose now that we have a pure solvent on side (1) and a solution on side (2) of semipermeable membrane. Pressure $p_2 > p_1$ such that the system is in equilibrium.

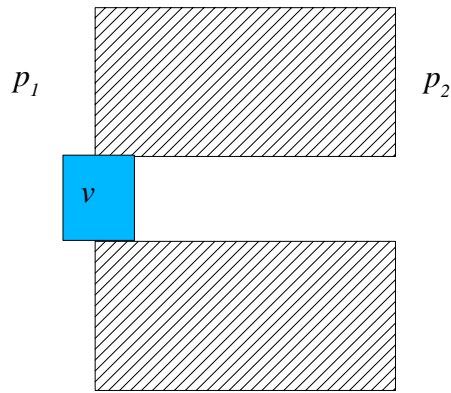

*Figure 7*

If a small volume *v* of water is transferred from side 1 to side 2 while the pressures $p_1$ and $p_2$ are kept constant, then at equilibrium the Gibbs function $G = U + pV - TS$ should remain constant too. The change in *G* is written as

(1) $\quad 0 = \Delta G = \Delta U + v \Delta p - T \Delta S$

Here

(2) $\quad \Delta U = U_2^v - U_1^v$

is the difference between the changes in internal energy when adding the a small volume *v* of water to side 2 resp. side 1 at constant pressures; similarly for *ΔS*. Finally we have written $\Delta p = p_2 - p_1$. As in the previous section we assume that the main contribution to *ΔS* comes from the diluting of the solute (see equ (8-7)), this gives us

(3) $\quad \Delta S = n k \dfrac{v}{V_2} = k c v$

where *n* is the number of solute molecules in volume $V_2$ and *c* is the solute concentration. Apparently we are led again to the van't Hoff formula if we may assume that *ΔU* = 0 in (1). We can reason that the concentration of water molecules on both sides at equilibrium are related by the Boltzmann factor (according to the microcanonical ensemble interpretation),

(4) $$\frac{c_1}{c_2} = e^{-\frac{\Delta u}{kT}}$$

where $\Delta u$ is the internal energy difference (2) calculated per molecule,

(5) $$\Delta u = \frac{\Delta U}{\rho \Delta V}$$

($\rho$ is the water concentration). Thus, $\Delta u \approx 0$ if we assume that $c_1 \approx c_2$ at equilibrium. This contrasts with "diffusion theory" which would assume that at equilibrium we would have a concentration gradient counterbalancing the hydrostatic pressure difference.

## 8.2 Thermodynamic explanation - the "classical version"

The usual treatment of the omsotic effect in thermodynamics text uses the concept of the chemical potential. The chemical potential is introduced in order to be able to handle systems with a varying number of particles. Thus, the 1st law of thermodynamics is rephrased as

(1) $$dU = -pdV + TdS + \sum_i \mu_i dN_i$$

The last term takes into account that the internal energy of the system may change because $dN_i$ particles of species #$i$ are added or removed (while volume $V$ and entropy $S$ remain constant). This energy $\mu_i$ per particle #$i$ is called its *chemical potential*. If we combine (1) with the definition $G = U + pV - TS$ of the Gibbs energy $G$ we obtain for its differential,

(2) $$dG = V dp - S dT + \sum_i \mu_i dN_i$$

Consider now the same situation as in section 8.1. At equilibrium a small volume of water is transferred from side 1 to side 2. The overall pressure and temperature are unchanged ($dp = 0$, $dT = 0$), so we obtain from (2),

$$0 = dG = dN \left( \mu^{(2)} - \mu^{(1)} \right)$$

that is (upper index refers to which side of the membrane the water is on),

(3) $$\mu^{(2)} = \mu^{(1)}$$

At equilibrium the chemical potential of water is the same on both sides of the membrane. This may seem quite "obvious" (though no mechanism is indicated), since if there were a difference in the chemical potential, water would spontaneously flow to the side with the smaller chemical potential as it would "pay off" energetically. Now, the chemical potential function of the water on the solution side differs from that of pure water due to two reasons: the higher pressure, and the presence of the solute. These two factors counteract such that the value of the chemical potential remains unchanged. We can thus write (3) as

(4)    $\mu(p,1) = \mu(p+\Delta p, \alpha)$

where α denotes the relative concentration of water,

(5)    $\alpha = \dfrac{c_{water}}{c_{water} + c_{solute}}$

The change in the chemical potential for relatively small changes in the pressure can be deduced using the relation

(6)    $\mu = \dfrac{\partial G}{\partial N}$

from which we get

(7)    $\dfrac{\partial \mu}{\partial p} = \dfrac{\partial^2 G}{\partial p \partial N} = \dfrac{\partial^2 G}{\partial N \partial p} = \dfrac{\partial V}{\partial N} = v$

noting that $V = \dfrac{\partial G}{\partial p}$ which also follows from (2). In (7) $v$ denotes the volume per water molecule. Thus, we may use the approximation

(8)    $\mu(p+\Delta p) \approx \mu(p) + v\Delta p$

Then, how does the chemical potential depend on the relative water concentration (5)? The traditional argument is based on assumption that we can use the ideal gas approximation for how the chemical potential varies with concentration. If we mix gases the *mixing entropy* is given by

(9)    $S = -k \sum_i N_i \ln(\alpha_i)$

where

(10)   $\alpha_i = \dfrac{N_i}{\sum_j N_j}$

is the relative concentration of species #$i$. Equ (9) describes the increase of entropy due to the increased total volume for the molecules to move in and can be derived using equ (8-7). As the expression does not depend neither on the pressure nor on the temperature it is argued that it can also be applied to the case of liquid (and even solid) mixtures. Using this approximation the chemical potential can then be written generally as

(11)   $\mu = \sum_i \mu_i(p,T) + kT \sum_i N_i \ln(\alpha_i)$

Using (11) we get for the change in the chemical potential due to the addition of *dN* water molecules to the solution

(12) $\quad d\mu = kT\, dN\, \ln(\alpha) = kT\, dN\, \ln(1-\beta) \approx -kT\, dN\, \beta$

where

(13) $\quad \beta = \dfrac{N_{solute}}{N_{solute}+N_{water}}$

is assumed to be small.

Combining this with (4) and (8) gives

(14) $\quad v\Delta p \approx kT\, \dfrac{N_{solute}}{N_{water}+N_{solute}}$

Since[31] $v(N_{water} + N_{solute})$ is about equal to the volume $V$ of the solution in the dilute case we arrive again at the van't Hoff formula for the osmotic pressure $\Pi = \Delta p$.

## 9. Statistical mechanics

Statistical mechanics provides a perspective on a level intermediary between thermodynamics and "molecular dynamics". Statistical mechanics is cognizant about the forms of the interactions between the molecules but does not try to solve the problem of describing the motions of individual molecules. In the previous section we already introduced some elements of statistical mechanics when using the Boltzmann formula (8.8). It is interesting to see how a more careful statistical mechanical treatment of the osmosis compares with the previous accounts. We may start with the "postulate" that for equilibrium the probability $p_i$ of a system, in contact with a thermal reservoir of temperature $T$, of being in a state $i\#$ with energy $E_i$ is proportional to the Boltzmann factor,

(1) $\quad p_i \propto e^{-\frac{E_i}{kT}}$

The entropy $S$

(2) $\quad S = -k \sum_i p_i \ln(p_i) = \langle -k \ln(p) \rangle$

and the internal (mean) energy $U$

(3) $\quad U = \sum_i E_i\, p_i = \langle E \rangle$

can[32] then be expressed by writing

---

31　The following assumption is equivalent to the assumption in 8.1 that water concentrations in terms of water molecules per unit volume are the same in pure water and the solution at equilibrium.

32　We use above the notation $\langle (...) \rangle$ here for the average $\langle x \rangle = \sum_i p_i x_i$ .

$$p_i = \frac{e^{-\beta E_i}}{Z(\beta)}$$

(4) $\quad Z(\beta) = \sum_i e^{-\beta E_i}$

$$\beta = \frac{1}{kT}$$

as

(5) $\quad U = -\frac{\partial \ln(Z)}{\partial \beta}$

and

(6) $\quad S = k\beta U + k\ln(Z)$

Thus, from the definition of the free energy $F$ (8.1) we get

(7) $\quad F = -\frac{1}{\beta}\ln(Z)$

$\quad\quad Z = e^{-\beta F}$

The energy of the water-solute system may be written as[33]

(8) $\quad H = H_0 + H_1$

where $H_0$ contains the water energy part, and $H_1$ contains the solute energy and the solute-water interaction energy. Thus we obtain in this case

(9) $\quad e^{-\beta F} = Z = \sum_{r,s} e^{-\beta H_0(r) - \beta H_1(r,s)} = \sum_s \left( \sum_r \left( e^{-\beta H_0(r) - \beta H_1(r,s)} \right) \right)$

where $r = (r_1, ...., r_N)$ refers to the positions for the water molecules, and $s$ similarly to the positions of the solute molecules. We neglect the kinetic energy part because we will be only interested in the dependence on (solution) volume $V$. Because we assume a low concentration for the solute we may assume there are no interactions between the solute molecules; i.e., we can neglect the solute-solute potential in (9). The parenthesis on the lhs of (9) can be written as

(10) $\quad \sum_r \left( e^{\beta H_0(r) - \beta H_1(r,s)} \right) = Z_0 \cdot \frac{\sum_r \left( e^{-\beta H_0(r) - \beta H_1(r,s)} \right)}{Z_0} = e^{-\beta F_0} \left\langle e^{-\beta H_1(r,s)} \right\rangle$

where

---

33 The following exposition parallels Ma (1985 ch. 18).

(11)  $$Z_0 = \sum_r e^{-\beta H_0(r)} = e^{-\beta F_0}$$

is the state sum for water, and the averaging <(...)> in the last term is the average over the water states. We suppose this average is approximately independent of the distribution *s* of the solute molecules and set

(12)  $$\sum_r e^{-\beta H_1(r,s)} = e^{-\beta W}$$

where *W* describes the effect of the solute-water interaction which depends on the temperature only, not on the (solution) volume *V*. Then, finally, taking the sum in (9) over all the configurations *s* of the solute molecules leads to the equation

(13)  $$e^{-\beta F} = e^{-\beta F_0 - \beta W} \left( \frac{V^n}{n!} \right)$$

We have introduced the combinatorial factor *n*! here though it is of no consequence in this case[34]. Thus, we obtain for the free energy

(14)  $$F = F_0 + W - \frac{1}{\beta}(n \ln(V) - n \ln(n) + n)$$

We have thus again a term $-n k T \ln(V)$ which gives the "osmotic pressure" formula e.g. using[35]

(15)  $$p = -\left( \frac{\partial F}{\partial V} \right)_T$$

If we also add the gravitational potential part, then for the minimum of the free energy we obtain,

(16)  $$0 = \left( \frac{\partial F_{grav}}{\partial V} \right)_T + \left( \frac{\partial F}{\partial V} \right)_T = p_{hydrostat} - p_{osmos}$$

Though this last statistical mechanical presentation of the osmotic effect does not give a "picture" how the process works it does make clear the assumptions that are needed for obtaining a valid approximation. For example, the strength of the water-solute does not play any significant role for the final expression (but it is important when finding the molecular mechanism for osmosis in each particular case). Indeed, from the intuitive point of view, if we had a very strong interaction between solute and water, then the water molecules hooked on to a solute molecule could be considered as part of the solute molecule.

---

34  For a system of *n* identical particles this factor ensures that the entropy (12) becomes extensive when *n* becomes large; i.e., the entropy of a combined system, $S(A \times B)$ will be equal to $S(A) + S(B)$.
35  This follows from $dF = d(U - TS) = dU - S\, dT - T\, dS = dU - S\, dT - (dU + p\, dV) = -S\, dT - p\, dV$.

Frank Borg ( borgbros@netti.fi ). Jyväskylä University, Chydenius Institute, PB 567, FIN-67101 Karleby.

## 10. Effective potential

The previous statistical treatment, though, may give a clue about how the osmosis works. The idea is to reverse the order of summation in equ (9.9). Summing over the solute molecule configurations we define an *effective potential correction* $U_{eff}$ for water (solvent) by,

(1) $$e^{-\beta U_{eff}(\mathbf{r})} = \sum_s e^{-\beta H_1(\mathbf{r},s)}$$

That is, by "integrating out" the solute contribution, the effect of the solute becomes expressed as a modification of the water potential function by an effective potential term,

(2) $$U_{eff}(\mathbf{r}_1,...,\mathbf{r}_N)$$

where $\mathbf{r}_i$ are the positions of the water molcules. Thus we have a potential difference between solvent water and pure water and this must be related to the "osmotic pressure".

## *Conclusion*

We have seen that the thermodynamic theory of osmosis relies only on the functional descritpion of the membrane (permeant for the solute and "invisible" for the solvent [water]), not on its mechanism and how the "nuts and bolts" of the membrane are constituted. Many of the "explanations" for osmosis try to explain osmosis in terms of a single mechanism, such as diffusion due to a presumed water-concentration gradient. None of the mehcnisms considered seem as such to be instrumental in making the osmosis happen. In our view, any mechanical description of osmosis has to go back to the very concept of pressure and its explication in terms of the virial theorem. This vindcates that solvent-water and pure water are indeed in different states and explains why pure water may flow into the solution. We suggest that the solute-retardation mechanism (section 4) could be used when looking for a simple but true explanation of osmosis in the textbooks when one wants to go beyond the thermostatic account. One important point in this context is to contrast the two cases of liquids on the one hand, and gases on the other hand. When the solute in the solvent is compared with the "ideal gas" it is only with respect to the fact that the solute molecules in a dilute solution do not interact noticeably. It is this circumstance which leads to the ideal-gas-like formula (1-1) for osmotic pressure. In thermodynamics this is a reflection of the fact that the chemical potential has the same form for liquids and gases. The virial theorem on the other hand derives the formula from the fact that there must be a solute-water interaction that cancels the gas-pressure that the otherwise free solute molecules would exert.

## *Appendix A. Demonstration of the virial theorem*

Consider particles that are restricted to a volume *V*. We are interested in the time average of the expression (the quantity $\sum_i \mathbf{r}_i \cdot \mathbf{p}_i$ was called the "virial" by Clausius)

(1) $$\frac{d}{dt}\left(\sum_i \mathbf{r}_i \cdot \mathbf{p}_i\right) = \sum_i \dot{\mathbf{r}}_i \cdot \mathbf{p}_i + \sum_i \mathbf{r}_i \cdot \dot{\mathbf{p}}_i = 2 E_{kin} + \sum_i \mathbf{r}_i \cdot \mathbf{F}_i$$

The sums are over all the particles, with locations denoted by $\mathbf{r}_i$, and momenta by $\mathbf{p}_i$, and on which act the forces $\mathbf{F}_i$. Since the particles are bounded by the finite volume *V* the time average of the lhs will approach zero,

(2) $$\left\langle \frac{d}{dt}\left(\sum_i \boldsymbol{r}_i \cdot \boldsymbol{p}_i\right)\right\rangle = \lim_{T\to\infty} \frac{1}{T}\int_0^T \frac{d}{dt}\left(\sum_i \boldsymbol{r}_i \cdot \boldsymbol{p}_i\right)dt \to 0$$

since $\sum_i \boldsymbol{r}_i \cdot \boldsymbol{p}_i$ remains bounded. It follows that ("virial theorem")

(3) $$0 = 2\langle E_{kin}\rangle + \left\langle \sum_i \boldsymbol{r}_i \cdot \boldsymbol{F}_i\right\rangle$$

In the last term the forces can be divided into two parts: the interparticle forces and the forces by which the walls of the vessel act upon the particles (and observed macroscopically as "pressure"),

(4) $$\sum_i \boldsymbol{r}_i \cdot \boldsymbol{F}_i = \sum_k \boldsymbol{r}_k \cdot \boldsymbol{F}_k^{(particle)} + \sum_j \boldsymbol{r}_j \cdot \boldsymbol{F}_j^{(wall)}$$

For the last term we obtain, taking the time average and going to the continuous limit,

(5) $$\left\langle \sum_j \boldsymbol{r}_j \cdot \boldsymbol{F}_j^{wall}\right\rangle = \int_{\partial V} \boldsymbol{r} \cdot \frac{d\boldsymbol{F}^{(wall)}}{dA}\,dA = -p\int_{\partial V} \boldsymbol{r}\cdot d\boldsymbol{A} = -p\int_V \nabla\cdot\boldsymbol{r}\,dV = -3pV$$

where the pressure $p$ is defined as the average force per unit area of the wall. Inserting this result back into the expression of the virial theorem we get,

(6) $$3pV = 2\langle E_{kin}\rangle + \left\langle \sum_k \boldsymbol{r}_k \cdot \boldsymbol{F}_k^{(part)}\right\rangle$$

The last term on the rhs can finally be put on the form

(7) $$\left\langle \sum_k \boldsymbol{r}_k \cdot \boldsymbol{F}_k^{(part)}\right\rangle = \frac{1}{2}\left\langle \sum_{j,k} \boldsymbol{F}_{j,k}^{(part)} \cdot (\boldsymbol{r}_j - \boldsymbol{r}_k)\right\rangle$$

where $\boldsymbol{F}_{j,k}^{(part)}$ is the force by which the particle at $\boldsymbol{r}_k$ acts upon the particle at $\boldsymbol{r}_j$.

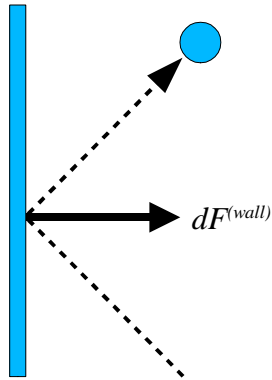

*Figure 8*

If we assume that the particles are distributed isotropically, such that the probability density $\rho^{(2)}(\boldsymbol{r}_1, \boldsymbol{r}_2)$ that a particle is found at $\boldsymbol{r}_1$ and another particle at $\boldsymbol{r}_2$, is of the form

(8)    $\rho^{(2)}(\boldsymbol{r}_1, \boldsymbol{r}_2) = \rho^2 g(|\boldsymbol{r}_1 - \boldsymbol{r}_2|)$

where ρ is the density, and that the interaction is independent of direction, then the virial theorem can be expressed as

(9)    $p = \rho kT - \rho^2 \frac{2\pi}{3} \int_0^\infty \frac{\partial u}{\partial r} g(r) r^3 dr$

Here $u(r)$ is the interaction potential and we have also used the equipartition theorem for setting the average kinetic energy to *kT/2* per (translational) degree of freedom.

## *Appendix B. Reverse osmotic experiment*

Hammel (2002) describes an interesting experiment where one first uses an osmometer with pure water on both sides in order to determine the hydraulic conductivity of the membrane. In the next experiment one adds the solute Dextran to one side and observes the ensuing osmotic flow. When the system has reached equilibrium (which corresponded to 10 cm water level difference), one removes part of the water from the pure water side which causes the water to start to flow in the reverse direction. It is found that the reverse flow is much smaller than the "forward" flow, the absolute total pressure difference, |*hydrostatic pressure - osmotic pressure*|, being the same. An obvious explanation for this seems to be that in the reverse flow the solute molecules are stopped at the membrane and thus exert a drag on the water. This drag can be simply estimated using the Einstein relation

(1)    $D = \frac{kT}{\gamma}$

which relates the diffusion constant $D$ of the solute to the friction constant γ of the solute molecule in water. Suppose water is flowing with velocity $v$ and that a proportion α of the *n* solute molecules have been stopped. Then their total drag force *f* per unit volume exerted on the water will be

$$(2) \quad f = \frac{\alpha n \gamma v}{V} = \frac{\alpha n k T v}{DV} = \frac{\alpha \Pi}{D} v$$

where $\Pi$ is the osmotic pressure given by the van't Hoff formula. For the forward flow an initial flow velocity of $v \sim 4.8$ cm/hr was reported. Assuming a diffusion constant of the order of $10^{-10}$ m$^2$/s we obtain for the factor $\Pi v/D \sim 10^8$ N/m$^3$ using $\Pi = 10$ cmH2O. We realize that the solute can exert a considerable drag on the water during reverse flow which may in fact be much larger than the flow resistance offered by the membrane. We obtain a very simple modell of the reverse flow if we assume the distribution of the solute to be uniform. Furthermore, the water level on the solution side is denoted $h_0$ and on the pure water side $h_1$. For reverse flow we have the condition

$$(3) \quad h_0 > h_1 + \frac{\pi_{osm}}{\rho g}$$

Now make a further simplifying assumption that at time $t$ the fraction of solute molecules participating in braking the water flow is

$$(4) \quad \alpha = \frac{h_0(0) - h_0(t)}{h_0(0)}$$

If the total mass of the liquid is $m$ ( $= \rho A l$ ) we can make the ansatz

$$(5) \quad m \ddot{h}_0(t) = -\left(1 - \frac{h_0(t)}{h_0(0)}\right) \frac{\pi_{osm}(0)}{D} \dot{h}_0(t) - \frac{A \rho}{L} \dot{h}_0(t) - A \rho g (2 h_0 - l) + A \left(\frac{h_0(0)}{h_0(t)}\right) \pi_{osm}(0)$$

Here $A$ is the cross-section of the tube, $l$ is the total length of the part of the tube filled with water, and $L$ is the hydrodynamic resistance of the membrane. The first term on the rhs describes the the solute drag, the second term is due to the membrane resistance, the third terms is due to the hydraulic pressure, and finally the last term is from the osmotic pressure taking into consideration that the solute concentration will increase as the volume $A h_0$ decreases. The lhs can generally be put to zero since accelerations can be assumed to die out quickly. Viscosity is neglected above (easy to add though) as it can be assumed that it is eclipsed by the membrane hydraulic resistance and the solute drag. The equ (5) makes of course no sense when $h_0(t)$ becomes very small because then the dilute solution assumption of the van't Hoff formula breaks down. The challenge is to develop a more detailed theory that can model the solute-membrane interaction, the non-homogenous distribution of the solute and describe how the "osmotic pressure" is affected in the non-homogenous case.

## Appendix C. Non-equilibrium thermodynamics

In the previous thermodynamic treatments of the osmotic effect we have studied global equilibrium states; that is, maximum entropy states (thermo-*dynamics* as thermo-*statics*). We clearly also need a theory about how these equilibrium states are reached via intermediary states as well as the time scales involved. Non-equilibrium thermodynamics does not offer a full dynamical picture of the phenomena but still goes a bit further than equilibrium thermodynamics. For osmosis it e.g.

prediccts a coupling between the solvent and the solute flow when the reflection coefficient is less than 1. The general theory uses balance equations and phenomenological laws in order to study small deviations from the equilibrium states. Consider as a simple example an iron bar that at a certain time has a non-uniform temperature distribution $T$. Left to itself the temperature of the iron bar will level out and reach a thermodynamic equilibrium state characterized by a uniform temperature. J Fourier proposed that the change of the temperature $T$ can be described by a differential equation (1-dimensional case):

(1) $$\frac{\partial T}{\partial t} = \kappa \frac{\partial^2 T}{\partial x^2}$$

This follows from an energy conservation equation

(2) $$\frac{\partial (\rho u)}{\partial t} + \nabla \cdot \mathbf{W} = 0$$

with internal thermal energy $u$ (per unit mass) given by ($c$ is the heat capacity per unit mass)

(3) $$du = c\, dT$$

and the heat flow given by

(4) $$\mathbf{W} = -\lambda \nabla T$$

As we have introduced a local internal energy function $u$, we might also propose a local entropy function $s$ (entropy per unit mass). The idea is that *locally* the system may be assumed to be in equilibrium and that it is therefore possible to define entropy locally even for a non-equilibrium system. Thus, employing local entropy $s$, and if we neglect the small volume changes, the change in entropy $ds$ will be related to the change in internal energy $du$ by $du = Tds$, and (2) can be written as an *entropy balance equation* (Sommerfeld 1952 p. 148),

(5) $$\frac{\partial (\rho s)}{\partial t} + \nabla \cdot \left(\frac{\mathbf{W}}{T}\right) = -\frac{1}{T^2} \mathbf{W} \cdot \nabla T$$

In the equation (5) we can interpret

(6) $$\frac{\mathbf{W}}{T}$$

as the *entropy flow* and

(7) $$-\frac{1}{T^2} \mathbf{W} \cdot \nabla T = \frac{\lambda}{T^2} |\nabla T|^2$$

as the *source* of entropy (local entropy production) which is apparently non-negative illustrating the law of increasing entropy for irreversible processes (spreading of the thermal energy in this case). The kind of reasoning leading to (5) can be generalized to cases involving e.g. hydrodynamic and electrochemical processes (Glansdorff and Prigogine 1971; Groot and Mazur 1986). The general starting point is the *Gibbs equation* (where we have added a term including $z_i F \Phi$ accounting for the energy of a charge with valence $z_i$ in a potential $\Phi$, $F$ = *Faraday* constant)

$$(8) \quad TdS = dU + p\,dV - \sum_i \left( \mu_i + z_i F \Phi \right) dn_i$$

together with mass conservation, energy conservation, momentum conservation, the equations of motion, and macroscopic phenomenological relations similar to the heat flow equation (4). One thus arrives at generalizations of the entropy balance equation (5)

$$(9) \quad \frac{\partial (\rho s)}{\partial t} + \nabla \cdot (\boldsymbol{\Theta}) = \sigma$$

Here $\boldsymbol{\theta}$ stands for the entropy flow, and $\sigma$ stands for the entropy source which generally may be expressed as

$$(10) \quad \sigma = \sum_k J_k X_k$$

where $J_k$ are named the *thermodynamic currents* and $X_k$ the *thermodynamic forces*. In the case (5) we have for instance

$$(11) \quad \begin{aligned} J &= \frac{W}{T} \\ X &= -\frac{1}{T} \nabla T \end{aligned}$$

If we consider a process that is close to the equilibrium (where $X = 0$) it might be assumed that the currents can be expressed as linear functions of the forces $X$,

$$(12) \quad J_k = \sum_j L_{kj} X_j$$

Equation (4) is an explicit example of (12), in the simple case of heat flow, with $L = \lambda$ (heat conductivity). In principle one could calculate the coefficients $L$ from the molecular theory[36] but usually one is happy to know them from experimental data. Also it turns out that sometimes knowing part of the coefficients one can predict other ones, since Onsager has proved the *Onsager reciprocal relations* (for a demonstration see e.g. Landau and Lifshitz 1980 § 120[37]),

$$(13) \quad L_{jk} = \pm L_{kj}$$

---

36 This problem can be related to *linear response theory* and the *Kubo-relations* derived by R Kubo (*J Phys Soc* (Japan) 12, 1957, 570). Another type of approach to calcluate the transport coefficients is based on *Boltzmann's kinetic equations* (see e.g. Reed and Gubbins 1973 § 12-8). The Boltzmann equations and their *Enskog approximations* are also discussed by Groot and Mazur (1984 ch. IX).

37 Finnish readers may consult the textbook by Arponen J and Honkonen J: *Statistinen Fysiikka* (3. ed. Liimes 2000) § 14.3. The result (13) follows from studying fluctuations near the equilibrium and taking into account the microscopic causality principle. For a mathematical discussion of the Onsager relations based on the theory of stochastic processes and the Fokker-Planck equations see Gardiner (2002 § 5.3.5).

The minus sign applies when the corresponding equations of motions for the forces involved are not time-symmetric (thus the minus sign applies e.g. for magnetic force). We might try to apply the above concepts to the osmotic flow through a membrane, considered as an irreversible process driven by a chemical potential difference $(\Delta\mu)_{p,T}$ (thermodynamic force $-(\Delta\mu)_{p,T}/T$) and a hydraulic pressure difference $\Delta p$ (thermodynamic force $-\Delta p/T$) over the membrane. After some algebra footwork and bookkeeping, and assuming a stationary flow (no accelerations), one obtains for the total entropy production (Groot and Mazur 1982 p. 435)[38]

$$(14) \quad \sigma_{tot} = \int_V \sigma \, dV = -j_1 \frac{(\Delta\mu_1)_{T,p}}{T} - j_2 \frac{(\Delta\mu_2)_{T,p}}{T} - j_v \frac{\Delta p}{T} = -\tilde{j}_1 \frac{(\Delta\mu_1)_{T,p}}{T} - j_v \frac{\Delta p}{T}$$

We may take component 1 to refer to the solvent, and the component 2 to refer to the solute. The last equality is obtained by using the Gibbs-Duhem relation

$$(15) \quad \sum \xi_k (\Delta\mu_k)_{p,T} = 0$$

and defining "volume flow" by

$$(16) \quad j_v = v_1 j_1 + v_2 j_2$$

and the "diffusion current" by

$$(17) \quad \tilde{j}_1 = j_1 - \frac{\xi_1}{\xi_2} j_2 \; .$$

What are the dissipative mechanisms accounting for the entropy production (14)? The pressure term can be associated with flow resistance (viscosity). The chemical potential term is associated with a concentration difference

$$(18) \quad (\Delta\mu_1)_{p,T} \approx \left(\frac{\partial \mu_1}{\partial \xi_1}\right)_{p,T} \Delta \xi_1 = \mu_{11} \Delta \xi_1$$

This concentration difference is not directly connected with a *diffusion of the solvent* through the membrane to the solute side, since, despite that we have a lower mass fraction $\xi_1$ of the solvent on the solute side, the concentration of the solvent in terms of *solvent molecules per unit volume* can in some cases be *higher* on the solute side than on the solvent side. Furthermore, the flow through the membrane may be in the form of a poiseuille flow in a narrow channel. The chemical potential term is rather associated with the *mixing entropy* as the additional solvent dilutes the solute.

It seems we have not yet learned anything new about the phenomenon of osmosis. However, we may use the ansatz (12):

---

[38] In the equations below the chemical potential is to be understood as epxressed not per mole but per molar mass. The currents $j_1$ and $j_2$ have then the he unit *mass/ time* and the specific volumes $v_1$ and $v_2$ have the unit *volume/mass*.

(19)
$$\tilde{j}_1 = -L_{11}\frac{(\Delta\mu_1)_{p,T}}{T} - L_{1v}\frac{\Delta p}{T}$$
$$j_v = -L_{v1}\frac{(\Delta\mu_1)_{p,T}}{T} - L_{vv}\frac{\Delta p}{T}$$

The coefficients $L_{ij}$ characterize the membrane and its interactions with the solvent and the solute, such as the hydraulic resistance. We define the reflection *coefficient a* by

(20) $$a = v_1\frac{\tilde{j}_1}{j_v} = u_1\frac{i_1 - i_2}{u_1 i_1 + u_2 i_2} \approx \frac{i_1 - i_2}{i_1}$$

where we have introduced (*m* is the total mass)

(21)
$$u_k = v_k \xi_k m \quad \text{(volume of species k)}$$
$$i_k = \frac{j_k}{\xi_k m} \quad \text{(specific current, mass per mass and time of species k)}$$

The approximate equality in (20) is valid for dilute solutions ($u_2 \ll u_1$). If the membrane is completely impermeable to the solute ($i_2 = 0$) we obtain $a = 1$, and if the membrane is indifferent to the solute as well as to the solvent we have $i_1 = i_2$ and we get $a = 0$. Next consider two cases. First, suppose we have a stationary state with $j_v = 0$, this yields according to (18)

(22) $$\frac{\Delta p}{\Delta \xi_1} = -\mu_{11}\frac{L_{v1}}{L_{vv}}$$

This defines the *osmotic pressure* $\Delta p$ which has to be applied in order to stop the flow. Secondly, consider the case where the flow is only due to a pressure difference $\Delta p$ ($\Delta\xi_1 = 0$); that is, a hydraulic flow, then we obtain from (18) and (20)

(23) $$\frac{\tilde{j}_1}{j_v} = \frac{L_{1v}}{L_{vv}} = \frac{a}{v_1}$$

Using the Onsager relation $L_{1v} = L_{v1}$ we obtain from (23), (22) and (20),

(24) $$\frac{\Delta p}{\Delta \xi_1} = -\frac{\mu_{11}}{v_1} a$$

If $a = 1$; that is, if the membrane is completely impermeable to the solute, then equ (24) will reduce to the usual van't Hoff formula if we use the ideal gas approximation for the chemical potential (8.2-11). From this we see that the reflection coefficient describes the proportion by which the real osmotic pressure differs from the van't Hoff pressure,

(25) $$\Pi = a \cdot \Pi_{vH}$$

Thus, using (23) we can write the first equation in (19) as

$$(26) \quad j_v = \frac{L_{vv}}{T}(a\Pi_{vH} - \Delta p)$$

The flow is proportional to the effective pressure difference, and the coefficient $L_{vv}/T$ can therefore be identified with the hydraulic resistance of the membrane. Solving (19) for the solute current and expressing it in terms of volume flow (26) and $\Delta c_2$ we obtain

$$(27) \quad j_2\left(v_2 + v_1\frac{\xi_1}{\xi_2}\right) = (1-a)j_v + \left(\frac{v_1^2 L_{11} - L_{vv}a^2}{T v_1}\right)\left(\frac{\partial \mu_1}{\partial c_2}\right)_{p,T} \Delta c_2$$

The second factor on lhs is in fact equal to $1/\rho_2$ where $\rho_2$ is the density of the solute (mass per unit volume). Hence we can finally write (27) as

$$(28) \quad j_2 = (1-a)\rho_2 j_v + P\Delta c_2$$

where $P$ is defined by

$$(29) \quad P = \rho_2\left(\frac{v_1^2 L_{11} - L_{vv}a^2}{T v_1}\right)\left(\frac{\partial \mu_1}{\partial c_2}\right)_{p,T}$$

The equations (26) and (28) are similar to the *Kedem-Katchalsky equations*[39] which can be written (Weiss 1996 p. 303) as

$$(30) \quad \begin{aligned} j_v &= L(a\Pi_{vH} - \Delta p) \\ j_2 &= \bar{\rho}_2(1-a)j_v + P\Delta c_2 \end{aligned}$$

Here $L$ is the hydraulic conductivity, $P$ is the permeability of the membrane, and

$$(31) \quad \bar{\rho}_2 = \frac{\rho_2(1) + \rho_2(2)}{2}$$

is the mean solute density in the membrane (taken as the average of solute density at either sides of the membrane). These equations can be easily extended to the multi-solute case by introducing reflection coefficients $a_k$ and permeability coefficients $P_k$ for each species. The physical implication of the Onsager relation is that it predicts the coupling (first term on the rhs in the second equation in (30)) between the solute flow and the solvent flow. For $a < 1$ the volume flow carries with it also solutes through the membrane. As described in § 5.4 Joos too suggested sort of a coupling between the solute flow and the solvent flow but this was to be effective also for the case $a = 1$.

---

[39] Kedem O and Katchalsky A (*Biochim. Biophys. Acta* 27, 1958, 229 - 256).

*Acknowledgments*

I am indebted to prof. Claus Montonen for finding the paper by Lorentz (1892) and sending me a copy. He also pointed out (and sent me a copy) that the later editions of Joos (1986) have a section on a kinetic theory for osmosis (this section is missing in the 1. edition). Prof. K Siegler was kind to send me a copy of the paper by Janáček and Siegler (2000). Prof. Gerald Pollack's witty remarks have forced me to think things over again which helped me to find some errors in my previous arguments. I am thankful to prof. Ted Hammel for a stimulating correspondence and providing me with a selection of his papers, photos and a preprint of his ms.

### *Readings*

(Note: Additional comments on older sources quipped from online antiquarian catalogues: see e.g. www.zvab.com. The highlighted items were not available at the time of writing.)

---

40 "Of signal interest, for it marks the beginning of the study of osmotic pressure..." (Fulton 150 ).

41 "The final results of Dutrochet's thirty years of research into plant and animal physiology, replacing a series of earlier papers that he considered 'null and void'. Dutrochet's most important contribution was the discovery of osmosis, to which he gave its name; he was the first to investigate the phenomenon systematically, and to recognize its fundamental significance in living organisms. He recognized that only plant cells containing green matter were capable of absorbing carbon dioxide, and was the first to detect the production of heat in both a plant and an insect muscle during activity" (Norman 673).

Frank Borg ( borgbros@netti.fi ). Jyväskylä University, Chydenius Institute, PB 567, FIN-67101 Karleby.

---

[42] "Includes an account of all the methods of determining osmotic pressure" (Garrison- M.). "Hartog J. Hamburger (1859-1924) was one of Donder's students, and became an assistant to Donders before assuming the chair of physiology at the University of Groningen in 1901. Hamburger's physico-chemical investigations covered topics such as permeability, hemolysis, and phagocytosis in which he made some notable contributions. Fascinated by van't Hoff's epochal work on osmotic pressure, he was the first to try to assess its significance for medicine and eventually published his findings in this comprehensive, three-volume work which includes an account of all methods of determining osmotic pressure. In the course of these studies, he disovered that bicarbotane and chloride ions could be exchanged across the cell membranes. Dr. chem. et med. H.J. Hamburger war Prof. der Physiol. an der Reichsuniversität Groningen, wo er eine bedeutende Schule gründete und 1911 ein neues, glänzend ausgestattetes Institut schuf. 'Hamburger, der bedeutendste physiolog. chemiker Hollands, förderte mit einer grossen Zahl grundlegender Untersuchungen, insb. auf physikal.-chem. Gebiete, sein Fach (Permeabilität der Zellen, Kalzium-Ionen bei der Phagozytose, Hämolyse, stereoisomere Zuckerarten, Vagus- und Sympathikusstoffe)' (Hirsch). Bedeutend für die Physiologie wurden u.a. seine Untersuchungen über das Verhalten roter Blutkörperchen in Salzlösungen verschiedener Konzentrationen, die Bestimmung der Physiologischen Salzlösung beim Frosch = 0,64% NaCl, Reistenzprüfung der Erythrozyten (1.Bd., S.378) und ferner seine Versuche über die Permeabilität tierischer Zellen gegenüber Salzlösungen. Garrison-M. 725; Hirsch Nz. 571; Astrup & Severinghaus, History of Blood Gases, p.161ff."

---

[43] This separate printing also contains: Van't Hoff, "Une propriété générale de la matière diluée" (pp. 42-49) and "Conditions électriques de l'équilibre chimique" (pp. 50-58). Garrison-Morton 706: "Van't Hoff stated that osmotic pressure is proportional to the concentration if the temperature remains invariable, and proportional to the absolute temperature if the concentration remains invariable." "To the analogy that exits between gases and solutions van't Hoff gave the general expression pV=iRT.. In "Lois de l'équilibre chimique dans l'état dilué, gazeux ou dissous" (1886) van't Hoff showed that for many substances the value of i was one, thus validating the relation pV=RT for osmotic pressure. It then became possible to calculate the osmotic pressure of a dissolved substance from its chemical formula and, conversely, the molecular weight of a substance from the osmotic pressure. In "Conditions électriques de l'équilibre chimique" (1886), van't Hoff gave a fundamental relation between the chemical equilibrium constant and the electromotive force (free energy) of a chemical process: ln K = -E/2T, in which K is the chemical equilibrium constant, E is the electromotive force of a reversible galvanic cell, and T is the absolute temperature" (D.S.B. 13: 579).

---

[44] "Marking the year 1877 was De Vries's paper entitled Untersuchungen it was a very valuable contribution to our knowledge of the subject. It contained a number of determinations of relative osmotic values of the different substances which are constituents of the cell sap, in the course of which De Vries established a general relationship between molecular weight and osmotic pressure. He also investigated the phenomenon to which he gave the name of plasmolysis, and showed how it can be applied to ascertain the osmotic pressure of various substances and solutions." (Green, J.R.: *History of Botany*, 1860-1900, pp. 253-254.)

Frank Borg ( borgbros@netti.fi ). Jyväskylä University, Chydenius Institute, PB 567, FIN-67101 Karleby.